\documentclass[11pt,a4paper,jmp,author-numerical]{revtex4-1}

\usepackage{graphicx}
\usepackage{amsmath}
\usepackage{amssymb,amsfonts}
\usepackage{bbold}
\usepackage{hyperref}
\usepackage{xargs}
\usepackage{mathtools}
\usepackage{tikz}
\usepackage{tikz-cd}
\usepackage{mathrsfs}
\usepackage{eucal}
\usepackage{physics}

\usepackage{xparse}
\NewDocumentCommand \deq { o m }{
\begin{equation}
#2
\IfNoValueF{#1}{\label{#1}}
\end{equation}
}

\def\Z{\mathbb{Z}}
\def\Q{\mathbb{Q}}
\def\R{\mathbb{R}}
\def\C{\mathbb{C}}
\def\FF#1{\mathbb{F}\!_{#1}}

\def\SU{SU}
\def\SO{SO}
\def\SL{SL}

\def\PGL{PGL}

\def\role{role}

\usepackage[T1]{fontenc}
\usepackage{natbib}

\usepackage{epigraph}
\usepackage[american,german,french]{babel}

\def\Laplace{\mathop{{}\bigtriangleup}\nolimits}

\DeclareMathOperator{\Conf}{Conf}
\DeclareMathOperator{\Gal}{Gal}
\DeclareMathOperator{\ord}{ord}
\DeclareMathOperator{\sgn}{sgn}
\def\AdS{H}

\numberwithin{subsubsection}{subsection}
\numberwithin{subsection}{section}
\numberwithin{equation}{section}

\begin{document}

\setlength\epigraphwidth{.6\textwidth}
\setlength\epigraphrule{0pt}
\renewcommand{\epigraphflush}{center}

\title{Holography and local fields}

\author{Ingmar Saberi}
\email{saberi@mathi.uni-heidelberg.de}
\affiliation{Mathematisches Institut \\ Ruprecht-Karls-Universit\"at Heidelberg}

\begin{abstract}
This paper, which is a summary (in which considerable creative license has been taken) of the author's talk at the sixth international conference on $p$-adic mathematical physics and its applications (CINVESTAV, Mexico City, October 2017), reviews some recent work connecting field theories defined on the $p$-adic numbers and ideas from the AdS/CFT correspondence. Some results are included, along with general discussion of the utility and interest of $p$-adic analogues of Lagrangian field theories, at least from the author's perspective. A few challenges, shortcomings, and ideas for future work are also discussed.
\end{abstract}

\maketitle

\section{Introduction}
\epigraph{{\it Das also war des Pudels Kern! \\
  Ein fahrender Skolast? Der Kasus macht mich lachen.}}{Goethe,  \fg{}Faust\og{}}

\selectlanguage{american}

Over the last couple of years, several papers have appeared~(\cite{Gubs1,HMSS,edgelength,O(N),GubserParikh,signs}; see also~\cite{bhattacharyya,ghoshal,osborne}), revisiting the topic of field theories on $p$-adic spacetimes in an attempt to develop an inherently discrete model of the physics of the AdS/CFT correspondence. Rather than going through calculations in any great detail, we would like to discuss the general arc of this work, emphasizing the analogies that are in play, the parts of the formalism that carry over straightforwardly due to these analogies, and the places where the analogy is incomplete and further work is necessary in order to be able to tell a fully satisfactory story. (For another perspective on the same material, see~\cite{Gubs-sum} from the proceedings of Strings 2016.)
The reader may not be intimately familiar with all of $p$-adic techniques, quantum field theory, and the ideas behind the AdS/CFT correspondence; indeed, the increasing dearth of generalists, even within the limited domain of mathematical physics itself, is likely responsible for the fact that these analogies (which are striking, once one knows where to look) went unnoticed for so long. For this reason, we will give at least a cursory and teleological review of each, in a way that we hope will make structural similarities apparent.

$p$-adic techniques have a long history in physics, and we will unfortunately not have the opportunity to offer a broader review of the literature here. See~\cite{Dragovich,DragovichKK} for recent reviews, and~\cite{BrekkeFreund} for a classic one. We will mention only that much attention was devoted to $p$-adic analogues of string theory~\cite[among others]{Ruelle,Volovich,FreundWitten,FreundOlson,Zabrodin,MarshakovZabrodin,RuelleTVW}, in which the string worldsheet (which, although it starts life as a two-dimensional smooth manifold, can be thought of due to some peculiarities of low-dimensional physics as an algebraic curve over~$\C$) was replaced by a $p$-adic algebraic curve. In this context, theories of scalar fields and analogues of conformal symmetry were studied, and objects such as Green's functions were first computed~\cite{Zabrodin,Melzer} The connection between $p$-adic models and hierarchical models in the style of Dyson---and therefore to renormalization group theory~\cite{Dyson,WilsonKogut}---was also remarked on some time ago; we will mention results of Lerner and Missarov~\cite{LernerMissarov,LernerMissarov2,Missarov,Missarov2}, among others.

\section{What is quantum field theory?}

We will begin by giving an acknowledgedly biased and incomplete definition of a quantum field theory. Of course, many such definitions have been given over the preceding decades, by people far more qualified to do so than the current author, and we do not pretend that our definition encompasses a complete and precisely delineated idea. Indeed, the term definition may be a misnomer, and perhaps a misleading one; let us instead say that we will offer a sketch of something called quantum field theory, without claiming that no other sketches are possible or useful.

We will say that a quantum field theory is a theory of
many identical local degrees of freedom,
parameterized by some geometric space~$X$,
and coupled together in a local and homogeneous way with respect to the geometrical structure of~$X$.

This is a definition that is broad enough to include many different things: standard Lagrangian field theories, of course, fit into it, as well as lattice field theories and even spin systems. And these are all systems that a physicist would think of as being described by quantum field theory, at least in some range of their parameters. 

To translate our sketch into more technical language, we might agree to say that 
a QFT is a quantum theory whose degrees of freedom are functions 
on~$X$. By a function, we will understand any locally defined object, such as a connection, tensor field, or a section of some other bundle; in the case when $X$ is a manifold, by the Serre--Swan theorem, these are just projective modules over the algebra $C^\infty(X)$, or another suitable algebra of functions. However, none of these beyond functions themselves will play a great {\role} in the story that follows.

Such a  function is understood to represent a measurement or observation that can be made independently everywhere in~$X$; one should imagine a quantity like the temperature or electric field in a room. We will always assume, in keeping with everyday experience, that measurements are made using real numbers, and therefore that fields are real- or complex-valued. More exotic number systems will appear in a different {\role}: namely, in the construction of different choices for $X$. 

We'll also imagine that 
the interactions of the theory are encoded in an \emph{action functional:} 
\deq{
S: \mathscr{F}(X) \to \mathbb{R}.
}
While it is perhaps not conceptually essential, the point of this requirement is to have a clear concept of the classical limit of the theories we are interested, and to be able to use pieces of the standard formalism (path integrals, Feynman diagrams, and the like) wholesale.

While we are of course being intentionally vague in our sketch, it should be emphasized that, at this point, $X$ could be anything: a manifold, of course, but also a lattice, a graph, or simply a set. Different such choices of $X$ will lead to different classes of models (lattice field theories, for example); these models will carry different amounts of structure, corresponding to the presence or absence of various geometric structures on~$X$ itself. In the case when $X$ is a set and no additional structures are present at all, we are just talking about many-body quantum mechanics.

One big question that has preoccupied researchers in the border regions of mathematical physics over the last thirty years or so might be stated as follows: How is geometric and topological information about $X$ reflected in the behavior of theories on $X$?
As is perhaps to be expected for such a broad question, many of the big developments of recent years can be interpreted as offering sketches of an answer. 

To just briefly mention one example, when $X$ is a smooth four-manifold, one theory that can be defined on it is a ``twist'' of $\mathscr{N}=2$ supersymmetric $\SU(2)$ gauge theory.
Considering this theory leads one to consider the moduli space $\mathscr{M}(X)$ of anti-self-dual instantons on~$X$. 
The work of Donaldson and others~\cite{Donaldson,Witten-TQFT}
showed that 
crude topological invariants (such as the homology) of~$\mathscr{M}(X)$ are \emph{sophisticated} invariants of the smooth structure of $X$ itself, which provide powerful tools for understanding four-dimensional smooth manifolds!

The use of field theories, in particular topological field theories, to construct and organize invariants of geometric and topological manifolds thus has a long and fruitful history. But it is natural to wonder if 
one can carry information across this bridge in reverse.
For some time, physicists have been speculating about the vague question of how the classical geometry of~$X$ is encoded in (e.g.) the Hilbert space of a theory, or the entanglement structure of states therein.

An idea for how to start sketching partial answers to such a question is as follows:  Vary $X$, so that different amounts of its structure are present in different ways, and use this to try and pick apart the question into smaller and more manageable pieces. From the author's viewpoint, this is nothing more than reasoning by analogy, which is perhaps the basic form of reasoning in all of pure math and theoretical science; we will expand a bit on the {\role} of analogy in the section that follows. After this, we will return to the question of where one might find interesting choices of~$X$ to think about.

\section{Analogy in mathematics}


In some sense, the whole story of mathematics (and of theoretical reasoning more generally) is a story about analogies. Much of the axiomatization that has taken place is about codifying analogies between different situations: when we say that the integers are an abelian group, we are calling upon an analogy that says that says that the integers are exactly like the real numbers, the integers modulo $n$, or the complex numbers, at least as far as properties that only have to do with the existence of a commutative composition law with inverses are concerned. The definition of ``an abelian group,'' then, is really just an embodiment of this particular analogy or shared piece of structure, so that its consequences can be explored in and of themselves, without reference to the particularities of any example.

Reasoning by analogy is of crucial importance in theoretical science too, being visible (just for example) in the simple but powerful idea 
that the same equations have the same solutions. (The importance of this point was hammered home to the author by teachers, many years ago.)  That is, when a model of a situation leads to a particular set of equations, precisely what is being modeled (along with what other equations might or might not be valid) becomes irrelevant, at least as far as properties depending only on those equations are concerned.


So far, we've just emphasized that structural reasoning, or abstraction, is, at its heart, a form of reasoning by analogy. But there are  examples of looser or larger-scale analogies in both physics and mathematics as well. 
For the former, analogies between quantum field theory and statistical mechanics come to mind; as to the latter, perhaps the best and most fruitful example of such a large-scale conceptual analogy is the 
function field analogy in number theory~\cite[for example]{Weil-genus,Poonen}. With an eye towards motivating the value of such analogies, we will recall (in brief and inexpert fashion) some aspects of it here. 

One route of approach into this analogy might proceed as follows: One is a number theorist, and as such is interested in studying integers. One of the key features of integers is that they admit a \emph{unique factorization} into prime numbers, up to an ambiguity of sign:
\deq{
n = u \prod_i p_i, \quad n \in \Z,
}
where the~$p_i$ are (not necessarily distinct) prime numbers, taken to be positive, and $u$ is a sign ($\pm 1$), or in other words a unit of~$\Z$.

Then, one starts to ask naive questions about the primes: How many primes are there of a given size, i.e., less than or equal to some integer $N$? What is the longest possible span of numbers that contains no primes? How close together can two consecutive large primes be? As everyone knows, many questions like these, while they are easy to state, are incredibly difficult to answer; some remain open to this day.

One way to think about difficult problems in number theory, then, is to look for \emph{toy models} exhibiting the same kinds of structural behavior---unique factorization, for instance---in a simpler environment. And one such model consists of polynomials; any polynomial over a field~$k$ can be factored into a product of irreducible components,
\deq{
f(x) = u \prod_i p_i(x), \quad f \in k[x],
}
where $p_i(x)$ are irreducible polynomials (taken to be monic), and $u$ is a unit (i.e.\ element of the ground field $k$). 

One now has a model where (unlike the integers) there are some parameters that can be tweaked: in particular, any choice of field will give a ring of polynomials with unique factorization, where similar questions can perhaps be formulated. Depending on the choice of field, understanding the behavior of polynomials may be more or less straightforward---and may have more or less in common with the behavior of integers.
For example, when the ground field $k=\C$, the irreducible polynomials are exactly the linear polynomials $x-t$ for $t\in\C$. 
This case is quite simple (the irreducibles are simple to understand completely), but also unlike the integers (for example, in that there are uncountably many irreducibles).

A big part of the utility of this analogy is the geometric picture it allows one to develop. For example, we have said that the integers can be thought of as being like regular polynomial functions on the complex plane (or, more generally, the affine line). The field~$\Q$ is then like the field of rational functions on the same curve. Moreover, a finite extension of~$\Q$ (or an algebraic number field) should then be analogous to a finite extension of~$k(t)$. But that is the same as a function field of some other curve $C$ over the ground field~$k$ (or possibly a finite extension of~$k$; see e.g.~\cite{CF, Cassels} for precise structure theorems on extensions of nonarchimedean local fields). Considerations like this lead one, for example, to the notion of the \emph{genus} of a number field, which was first defined by Weil~\cite{Weil-genus}.

Now, each irreducible polynomial (which one can think of as closed points of the curve $C$, or equivalently as $\Gal(\bar{k}/k)$-orbits of $\bar{k}$-points of~$C$) defines a valuation on the function field, roughly speaking by order of vanishing at the corresponding point. This is perhaps easiest to think about when $C$ is a Riemann surface, and the corresponding valuation is literally the order of pole or zero of a rational function at a point $p\in C$. Of course, there are also valuations corresponding to the points at infinity of $C$, i.e., to points in its projective closure~$\overline{C}$; in the case of the affine line, the valuation at infinity is just the valuation by the degree of the polynomial. By well-known theorems, perhaps most familiar from complex analysis, not all of these valuations are independent: in fact, they satisfy the single relation 
\deq{
\sum_{p \in \overline{C}} \ord_p(f) = 0,
}
for any rational function $f$ on~$C$. 

One can complete the field with respect to any of its valuations. For simplicity, let's just think about the origin inside of the affine line. The completion of $k[t]$ with respect to the corresponding valuation (which is just by powers of~$t$) is, of course, the ring of formal power series $k[\![t]\!]$, and the corresponding completed field is the field of formal Laurent series $k(\!(t)\!)$. It is fruitful to think of this as looking at functions ``locally'' around a point of~$C$, 

The best analogy is obtained by letting $k = \FF{q}$, the finite field with $q=p^n$ elements. 
In fact, by results of Artin and Whaples~\cite{Artin-Whaples}, this gives a complete list of all ``global'' fields: they are either number fields (finite extensions of~$\Q$) or function fields of curves over a finite base field~$k$. A global field is defined by having a set of places or valuations, satisfying a relation, and for which at least one valuation is either Archimedean or discrete with finite residue field.
``Local'' fields are then completions of global fields at a chosen place; equivalently, they are locally compact topological fields, with respect to a non-discrete topology.

A key way in which the analogy is strengthened when $k=\FF{q}$ is that all the prime ideals of the ring of integers are then of finite index, and there are finitely many primes of a given polynomial degree, just as for~$\Z$. So it is meaningful to ``count'' primes and ask questions about their density. A sharp analogue of the Riemann hypothesis for function fields can be given, in the form of a structure theorem about the Hasse-Weil zeta function of the curve~$C$; it was conjectured by Artin and proved by Weil (and earlier by Hasse for elliptic curves). Similar statements for other varieties, the famous  Weil conjectures later proved by Deligne, can thus be thought of as the fruit of this analogy; unfortunately, though, this is as far as we can go into this fascinating topic here.

As already alluded to, $\Q$ is also a global field, and so has a relation between its different valuations. There is one of these corresponding to each prime $p$ (analogous to  the valuations measuring order of vanishing at a closed point); these are simply defined by
\deq{
x = p^\nu \frac{a}{b} \quad \Rightarrow \ord_p(x) = \nu, \quad |x|_p = p^{-\nu},
}
for $a$ and $b$ prime to~$p$. There is also the usual notion of absolute value on~$\Q$; a theorem of Ostrowski says that, up to equivalence, these are the only nontrivial valuations. Moreover, they satisfy the obvious relation
\deq{
\prod_p |x|_p = 1,
}
where $p$ runs over all places, i.e., the primes together with infinity.


The locally compact completions of~$\Q$ are therefore the real numbers and the $p$-adic fields; these and their finite extensions exhaust the locally compact topological fields of characteristic zero, and they will be important characters in the rest of our story. Based on the function field analogy, one can think of the $p$-adics as being like ``germs of functions around the point $p$;'' however, it is important to note that none of these local fields are isomorphic.

We will not take the time to review many of the peculiar properties of the $p$-adic norm here; we will content ourselves with giving just one piece of intuition. The $p$-adic norm (as well as any valuation coming from a similar concept, such as order of vanishing or divisibility) is best understood in terms of a hierarchical structure, like the successive division of species into smaller and smaller taxa. The closeness of two elements is then understood in terms of the lowest ``height'' of taxon at which they fall into the same group. 

\section{Structural ingredients of a Lagrangian field theory}


To come up with something that fits our sketch of a field theory, it's probably a basic requirement to be able to make sense of the concepts of \emph{locality}, \emph{homogeneity}, and \emph{isotropy}.

Locality suggests that
$X$ has a notion of \emph{distance}, \emph{measure}, or \emph{causal structure}, which is respected by the interactions in the theory.
In a standard field theory on affine space (or a smooth manifold), 
 this often means something like
\deq{
S[\phi] = \int_X \mathscr{L}[\phi(x)].
}
So we need a notion of a ``local Lagrangian density'' (a map $\mathscr{L}$ from functions to functions, densities, or top forms, depending only on functions and their derivatives at each point), as well as a notion of integration that assigns a number to a local Lagrangian density. 

For lattice models, though, we might need a different notion of local density: it's too strong to require that individual lattice sites interact only with themselves. Instead, we would ask for something like ``nearest-neighbor'' interactions: i.e., $S$ is a sum of terms corresponding to each site, where the term associated to a site depends only on the degrees of freedom from some finite neighborhood of that site. 
The notion of locality might even be further relaxed, to mean that interactions between separated points decay sufficiently fast (exponentially, for example) with distance.

Now, depending on the choice of $X$, there may be ``spacetime'' symmetries (implemented by the action of a group $G$ on~$X$). The theory may or may not respect the action of these symmetries on the fields~$\mathscr{F}(X)$.

Often (when $X$ is a maximally symmetric space), some group acts freely transitively on~$X$. The theory is \emph{homogeneous} when it respects the action of such a spacetime symmetry, which plays the {\role} of translations (and in fact consists of translations, in the case when $X$ is just affine $\R^n$).

But the group of translations may not be the whole story. For $X=\R^n$, the group of affine transformations is a semidirect product of translations and rigid rotations, 
\deq{
G = \R^n \rtimes \SO(n).
}
Poincar\'e-invariant theories always preserve these two symmetries, as makes sense for any theory that is formulated to depend on the metric, but not on any specific choice of coordinates. Translation invariance represents homogeneity, and rotation invariance represents \emph{isotropy} (or, in Lorentzian signature, Lorentz invariance); in other cases, the combined notion of a theory invariant under isometries of~$X$ will make sense, even when the splitting as a semidirect product does not.

But there are other symmetries present, which may or may not be preserved in interesting ways, and may respect less structure than Poincar\'e transformations.
For example, there are the 
\emph{discrete symmetries} of quantum field theory (in particular, $P$ and $T$), coming from the group of components of the Lorentz group. These may or may not be respected, and are in fact broken in the standard model of particle physics.

There are also symmetries that act on the spacetime, while failing to respect the metric structure in prescribed ways. An important symmetry is \emph{scale invariance}. While most quantum field theories of interest fail to be scale-invariant, the study  of how scale invariance is broken is the theory of the renormalization group.

One generally expects local scale-invariant theories to be invariant under a larger group of transformations: the 
 \emph{conformal group}, consisting of local (position-dependent) scale transformations. (While general reasoning about locality, similar to the reasoning that accompanies the promotion of global to gauge symmetries, suggests that scale-invariant theories are often conformal, there are subtleties related to this idea; we do not go into these here. For a recent review, see~\cite{Nakayama}).

In order to be able to write down something that deserves to be called a standard field theory Lagrangian, we will need a couple of additional structures. In particular, we would like to be able to make sense of a standard-looking kinetic term, which (in momentum space) looks something like this:
\deq{
S[\phi] = \int_{X^\vee} \phi(-k) \left( |k|^2+m^2 \right) \phi(k) + \cdots
}
Of course, making sense of this expression just means being able to assign a meaning to each of the symbols. 
Let's think about what that entails.

First off, the dual space  $X^\vee$ has to be defined. That means $X$ has a notion of (mutually commuting) translation symmetries, with a complete basis of eigenfunctions $\phi_k \in \mathscr{F}(X)$, diagonalizing those translations. Here the parameter~$k$ takes values in $X^\vee$, which is by definition the joint spectrum of the translation operators.
This amounts to saying that there is a notion of mode expansion, or equivalently, of the Fourier transform. (Of course, this requirement isn't absolutely necessary---of course, such a requirement fails for field theories on generic curved spacetimes---but we'll keep it around for now.)

In the case of~$\R^n$, we further have that $X \cong X^\vee$. But this isn't necessarily true: in lattice models, for example, the relevant fact is $\Z^\vee = S^1$. Physicists would refer to this under the name of the ``Brillouin zone.''

I will also need a notion of ``size'' on~$X^\vee$ (generalizing the length of a vector). And, finally, I'll need a notion of integration on~$X^\vee$.
Luckily, one set of assumptions is enough to ensure both a notion of translation-invariant integral and a good notion of Fourier transform. Whenever the translations of~$X$ are a locally compact abelian group, one can integrate with respect to Haar measure, and the Pontryagin duality theorem ensures that we have a good theory of the Fourier transform.

Once I can do this, I can make sense of the kinetic term of a real scalar field, and so I'm really in familiar territory.
The essential point in this is to identify the structures that allow us to make sense of such a term. The more of these structures $X$ has, the closer one is to being able to define a theory on it in perfect analogy to your favorite typical QFT.

\section{Parallelism}

We've catalogued many of the important structures on~$\R^n$ by now. To reiterate, the central idea is that many of these structures exist just because $\R^n$ is  an affine space over a locally compact field. 
Taking our cue from mathematics, we'll therefore try to define field theory models over other locally compact fields, and see how far we can get. 
We will hope to rely on the fact that, at least as far as the algebraic structures are concerned, affine spaces over fields all behave similarly.

We have already recalled that the $p$-adics are the completions of~$\Q$ at its finite places. A generic element of~$\Q_p$ can be written in the form
\deq{
 \sum_{i = -\nu}^\infty a_i p^i,
 }
 where the digits $a_i \in \{0,\ldots,p-1\}$; the corresponding sequence of partial sums is always Cauchy with respect to the $p$-adic distance, and thus by definition has a limit in the completion.
 They are locally compact fields (with respect to a nondiscrete topology), and therefore have additively and multiplicatively invariant Haar measures; we will denote the additive-invariant measure simply by~$dx$. The multiplicative Haar measure is then simply $dx/|x|$.

The theory of additive characters of~$\Q_p$ is completely understood: they are of the form
\deq[achars]{
\chi_p(kx) = e^{2\pi i \{k x\}_p},
}
where $\{ x \}_p$ denotes the $p$-adic ``fractional part:'' it is the natural map 
\deq{
\Q_p \rightarrow \Q_p/\Z_p \subset \Q, \quad
 \sum_{i = -\nu}^\infty a_i p^i \mapsto \sum_{i=-\nu}^{-1} a_i p^i. 
}
For the sake of unity, we'll let $\{x\}_\infty = x$ for real numbers. Now, additive characters span the space of well-behaved (locally constant) functions on~$\Q_p$, so that there is a good theory of the Fourier transform. \eqref{achars} exhibits the fact that~$\Q_p$ is self-dual.




Many of the classical special functions of mathematics can be constructed using these ingredients. For example, Tate's famous thesis~\cite{CF} contains (among many other things) a recipe for constructing the local zeta function of any locally compact, Fourier-self-dual field. Let $\theta(x)$ be the fixed point of the Fourier transform. One should therefore think of it as an analogue of the Gaussian; 
in fact,
\deq{
\theta_\infty(x) = e^{-\pi x^2},
}
while $\theta_p(x)$ is the characteristic of the unit ball ($p$-adic integers). 
\deq{
\int \frac{dx}{|x|} \,\theta(x) |x|^s = \zeta(s)
}
It is then a good exercise to check that 
\deq{
\zeta_p(s) = \frac{1}{1-p^{-s}}, \quad \zeta_\infty(s) = \pi^{-s/2} \Gamma(s/2),
}
so that Tate's definition recovers precisely the local factors of the ``completed'' Riemann zeta function. We refer the reader to e.g.~\cite{Gelfand} for a more thorough discussion of $p$-adic Lie groups, special functions, and character theory.

There are natural pseudodifferential operators, most easily defined in momentum space as multiplication by~$|k|^s$. (Here $s$ is a parameter controlling the ``order'' of derivative, which may take on real values; these operators, which are nonlocal in position space for $p$-adic fields, are the \emph{Vladimirov derivatives}. For a quick review of some of their properties, see the appendix of~\cite{edgelength}; a more complete reference is~\cite{Vladimirov-book}.)







As all of the preceding discussion hopefully suggests, $p$-adic analogues of familiar field theory models, such as the $O(N)$ model, can now be defined straightforwardly. The resulting models correspond, in the structural way that we have outlined, to the ordinary models; in some sense, they can be thought of as somehow intermediate between standard continuum field theories and lattice models. For example, the $p$-adic line (unlike the lattice) admits a scaling symmetry, but that symmetry is discrete rather than continuous. There are even some novel features: for instance, while marking an open set or interval in the real line breaks its translation symmetry, there exists a nontrivial subgroup of the $p$-adic translations preserving any chosen open set. 

 The utility of such models has recurred across mathematics;
in general, one can think of a hierarchical (or $p$-adic) model as replacing the real number line by an analogue, that has even more powerful geometric and algebraic structures---at the expense of its one-dimensional structures (ordering, path connectedness, \dots)~\cite{Tao}. See also~\cite{Manin1991} for related discussion.

Computations in theories defined at general places (i.e., over the reals or over the $p$-adics) sometimes exhibit universal answers, independent of which place the theory is defined at! The independence of the result on the place, however, has to be suitably interpreted. What seems to happen is that the result can be expressed in terms of special functions, which admit a uniform definition at all places. The numerical values of the results thus do not necessarily agree---but the structural form of the results is identical.

The example of this I have in mind is a result, due to Gubser, Jepsen, Parikh, and Trundy, for leading-order anomalous dimensions in the $O(N)$ model~\cite{O(N)}. 
The model is defined by a standard-looking action:
\begin{multline}
S = \int dk\, \phi^i(-k) \qty( |k|^s + m^2 ) \phi^i(k) \\ + \int dk_1\cdots dk_4\, \frac{\lambda}{4!} T_{i_1i_2i_3i_4} \phi^{i_1}(k_1)\phi^{i_2}(k_2)\phi^{i_3}(k_3)\phi^{i_4}(k_4) \, \delta\qty( \sum_i k_i).
\end{multline}
The tensor $T$ just contains the three possible contractions of indices,
\deq{
T_{i_1i_2i_3i_4} = \frac{1}{3} \qty( \delta_{i_1i_2} \delta_{i_3i_4} + \delta_{i_1i_3} \delta_{i_2i_4} + \delta_{i_1i_4} \delta_{i_2i_3} )
}
The anomalous dimensions
for the operators~$\phi$ and~$\phi^2$ in this model are given by
\deq{
\gamma_{\phi,\phi^2} = \Res_{\delta=0} g_{\phi,\phi^2}(\delta) + O(1/N^2)
}
where the functions $g_{\phi,\phi^2}$ are given by
\deq{
\begin{split}
g_\phi(\delta) &= \frac{1}{N} \frac{B(n-s,\delta-s)}{B(n-s,n-s)}, \\
g_{\phi^2}(\delta) &= -\frac{2}{N} \frac{B(n-s,\delta-s)}{B(n-s,n-s)} + \frac{1}{N}\frac{B(\delta,\delta)}{B(n-s,n-s)} \left( 2 \frac{B(n-s,n-2s)}{B(n-s,n-s)} - 1 \right).
\end{split}
}
Note that these results apply equally well to real or to $p$-adic models, assuming the special functions involved are defined uniformly! The gamma and beta functions that appear are defined (for a theory on $\R^n$ or $\Q_p^n$, which here means the unique unramified degree-$n$ extensions of $\Q_p$) by the relations
\deq{
 \Gamma_p (s) = \frac{\zeta_p(s)}{\zeta_p(n-s)}, \quad B_p(t_1,t_2) = \frac{\Gamma_p(t_1)\Gamma_p(t_2)}{\Gamma_p(t_1+t_2)},
}
$p$, as usual, is a place (a prime or infinity).


\section{Exactly solvable fermionic models}

As another example of an interesting model that can be treated uniformly across places (and exhibits some interesting new features in the non-Archimedean case), one can consider models of interacting fermions on the line~\cite{signs}, 
 defined in analogy to the recently introduced Klebanov--Tarnopolsky models. 
 

To define fermionic models on the $p$-adic line, one needs to
 use Grassmann field variables. Of course, this requires that the
 the symmetric quadratic form (propagator) in the kinetic term be replaced with an antisymmetric one.
 Normally, this is achieved by writing an action with one time derivative, instead of two; however, since any order of Vladimirov derivative defines a symmetric form, a little more thought is required.

In the real case, the propagator of a free fermion is 
\deq{
G(k) \sim \frac{1}{k} = \frac{\sgn(k)}{|k|}.
}
So, by analogy, one can antisymmetrize the propagator by using a quadratic multiplicative character of the field.
It's also possible to introduce more than one flavor of fermion, and to contract flavor indices with an antisymmetric tensor.

This leads one to the following class of actions for theories of interacting fermions:
\deq{
\begin{split}
S_\text{free} &= \int d\omega\, \frac{1}{2}\phi^{abc}(-\omega)\, |\omega|_p^s \sgn(\omega)\, \phi^{abc}(\omega), \\
 S_\text{int} &= \int dt\, \phi^{abc} \phi^{ab'c'} \phi^{a'bc'} \phi^{a'b'c}.
 \end{split}
}

Here, the field is either commuting or anticommuting; pairs of flavor indices are contracted either with $\delta$ or with a fixed antisymmetric matrix~$\Omega$; and the sign character may be either  ``even'' or ``odd,'' meaning that $\sgn(-1)=\pm1$.
There is already a constraint on these choices: for the kinetic term to be nontrivial, we must have  that $\sigma_\psi \sigma_\Omega = \sgn(-1)$.
In fact, a second constraint appears in the infrared, so that exactly one specific collection of these choices leads to consistent behavior at each place! In addition, to ensure that the field has positive scaling dimension and that the interaction term we write is relevant, we ask that the spectral parameter satisfy ${1/2}<s\leq 1 $.

Just like the ordinary Klebanov--Tarnopolsky model (and other models of SYK type), this theory is dominated in the large-$N$ limit by the ``melon'' diagrams.
These melon diagrams can be resummed into an exact Schwinger--Dyson equation, which determines the two-point function in the interacting large-$N$ theory. 
In the limit of large~$N$ and weak coupling, 
with $g^2N^3$ held fixed, this Schwinger--Dyson equation takes the form
\deq{
G = F + \sigma_\Omega\, (g^2N^3)\, G\star G^3 \star F.
}
Here $G$ is the interacting, and $F$ the free, two-point function.
In the infrared, this equation can be solved, giving a universal limiting behavior:
\deq{
G(t) =b  \frac{\sgn(t)}{|t|^{1/2}},\quad |t| \gg (g^2N^3)^{1/(2-4s)}
}
where
\deq[eq:norm]{
 \frac{1}{b^4g^2N^3} = -\sigma_\Omega \Gamma(\pi_{-1/2,\sgn}) \Gamma(\pi_{1/2,\sgn}).
}
Notice that the scaling in the IR limit is completely independent of the spectral parameter of the UV theory!

Here we are making use of a generalized gamma function, depending on a multiplicative character of the field:
\deq{
\Gamma(\pi) = \int \frac{dt}{|t|} \, \chi(t) \pi(t).
}
The characters appearing in~\eqref{eq:norm} are just products of characters coming from the norm and from the specific quadratic character appearing in the propagator, i.e.,
\deq{
\pi_{s,\sgn}(t) = |t|^s \sgn(t).
}

{
For fermionic theories with direction-dependent characters (i.e., characters that are nontrivial when restricted to~$\Z_p^\times$), one can do even better. It is possible to explicitly solve the Schwinger-Dyson equation for behavior at all scales, interpolating between the UV and the universal IR!
If we introduce the notation that $F(t) = f(|t|) \sgn(t)$ (and similarly for~$G$), then the Schwinger--Dyson equation reduces to
\deq[redSD]{
g = f + \sigma_\psi \, \frac{
g^2 N^3}{p} |t|^{2} g^4 f.
}
And this quartic can be explicitly solved for~$g$ when $\sigma_\psi = -1$. However, when~$\sigma_\psi=+1$ (i.e., the fields are bosonic), \eqref{redSD} does not have a well-defined limit in the deep infrared. 
As such, while fermionic theories have a well-defined two-point function at all scales, the IR limit of theories with bosonic fields appears to be problematic.

\section{Holography}

In all of what we've said so far, we've just discussed $p$-adic analogues of Poincar\'e-invariant field theory. We haven't said much about conformal transformations, or about the AdS/CFT correspondence, at all. But the original motivation behind recent papers revisiting $p$-adic field theory was to investigate whether the discrete scale symmetry present in such models could be used to construct analogues of AdS/CFT that naturally live on discrete bulk spaces. We will say a few words about this now, beginning with a brief reminder of some of the central ideas of AdS/CFT.




Let $\R^{n,m}$ denote an $(n+m)$-dimensional real space, equipped with the standard metric of signature $(n,m)$. $m$ will be zero for Euclidean signature, and one for Lorentzian; when it does not appear, it is taken to be zero.
The group of conformal transformations of~$\R^{n,m}$ is $\SO(n+1,m+1)$. And the anti-de~Sitter space~$\AdS$ of signature~$(n,m)$ can be constructed by taking the vectors of unit norm inside~$\R^{n,m+1}$; $\SO(n,m+1)$ naturally acts on this space, and it has a universal cover which is $\AdS^{n,m}$. For $m=0$, we are just constructing the standard hyperbolic space. Note also that $\AdS$ is a maximally symmetric space: it can be thought of as the quotient
\deq{
\AdS^n = \SO(n,1)/K,
}
where $K\cong \SO(n)$ is a maximal compact subgroup.

So the group $\Conf(n)=\SO(n+1,1)$ naturally appears acting on two different spaces: on~$\AdS^{n+1}$ by isometries, and on~$\R^{n}$ (or its compactification $S^n$) by conformal transformations. In fact, there is a good reason for this. While a metric on~$\AdS^{n+1}$ does not induce a metric on the boundary at infinity (because the metric diverges there), the conformal class of the induced metric \emph{is} well-defined. So isometries act on the asymptotic boundary $S^n$ in a way that preserves the conformal structure.

As we've emphasized, symmetries of a space $X$ can often be expected to act on a theory defined on~$X$. 
As such, there are two natural ways to make $\Conf(n)$-invariant (Euclidean) theories:
One can take $X=S^n$ (or $\R^n$), and look for conformal theories: fixed points of renormalization group flow.
Or, one can choose $X=H^{n+1}$, and take any field theory! 

It is then perhaps natural to wonder if a field theory on hyperbolic spacetime naturally couples to a (conformally invariant) field theory at asymptotic infinity. Another clue was given by the Bekenstein--Hawking formula for the entropy of a black hole,
\deq{
S = \frac{c^3}{4 \hbar G} A,
}
where~$A$ is the area of the event horizon. Speaking very loosely, one normally expects the entropy to be a count of the number of degrees of freedom, and so to be \emph{extensive:} that is, for a field theory, it should scale with the volume of a region. But Bekenstein and Hawking's results (along with other ideas in general relativity) were interpreted to suggest that the entropy in a theory of gravity scales with the area of the \emph{boundary} of the region. This is intuitively because of the presence of an enormous group of gauge symmetries (diffeomorphisms).

The AdS/CFT correspondence~\cite{GKP,Witten-AdS,Maldacena}, which grew out of observations like these, states that (certain) conformally invariant theories on~$S^n$ are \emph{equivalent} to gravitational theories on~$H^{n+1}$. The original construction of a pair of theories proposed to be dual in this fashion relied on string theory; the worldvolume theory of a large stack of coincident $D3$-branes (maximally supersymmetric Yang--Mills theory on~$\R^4$, in the large-color limit) is expected to be dual to the gravitational theory defined by type-IIB string theory on the near-horizon geometry of the branes, $\AdS^5\times S^5$.

But a precise understanding of the correspondence involves not just the prediction that two theories are equivalent, but a set of statements about how constituents of (and computations in) the two theories are meant to correspond to one another. The first few entries in this dictionary were worked out in bottom-up fashion by Witten~\cite{Witten-AdS}, who proposed the following ansatz:
\deq[ansatz]{
 \left\langle \exp \int \phi_0 \mathscr{O} \right\rangle_\text{CFT} = Z_\text{bulk}(\phi_0)
 \approx \exp \left(  -S_\text{cl}(\phi_0) \right) . 
 }
Here $\phi_0$ is a particular boundary condition at asymptotic infinity for the bulk field~$\phi$. The statement is thus that fields in the bulk correspond to operators in the boundary theory, with the coupling by taking the boundary value of the field as a smearing function for insertions of the corresponding operator in the CFT. The inverse transformation is possible because of a particular crucial property of the Klein--Gordon (or Laplace) equation on hyperbolic space: it has precisely one solution extending any chosen boundary condition at asymptotic infinity.
This makes the correspondence sensible in the limit where the bulk dynamics are classical.

The existence of a unique solution to the generalized Dirichlet problem for  the equations of motion can be expressed in terms of a bulk-to-boundary Green's function, which is the unique solution extending a delta-function boundary condition. This Green's function is a key object in AdS/CFT. In the half-space model of~$H^{n+1}$, i.e. $\R_+ \times \R^n$ with metric 
\deq{
ds^2 = \frac{1}{x_0^2} \sum_i dx_i^2,
}
such a Green's function for a massless field is explicitly given by
\deq{
 K(x) = \frac{x_0^n}{(x_0^2 + x_1^2 + \cdots + x_n^2)^n}.
 }

As one might expect, both fields and operators are classified by representations of the same group, $\Conf(d)$, and so the Casimir operators also correspond; in particular, the mass of the bulk field corresponds to the conformal dimension of the boundary operator.
The precise correspondence between these two quantities can be extracted from  the asymptotic behavior of~$K(x)$, by demanding that~\eqref{ansatz} make sense; the scaling dimension~$\kappa$ of the operator must therefore be
\deq{
 \kappa = \frac{1}{2} \left( d + \sqrt{d^2 + 4 m^2} \right).
 }
This also suggests the intuitive idea that the extra, radial coordinate of~$\AdS$ plays the role of a scale in the boundary theory.


Now, in low dimensions, due to exceptional isomorphisms, the whole story of conformal invariance can be formulated algebraically, without reference to a metric~\cite{ManinMarcolli}:
\deq{
\begin{split}
S^2 = P^1(\C), &\ \,H^3 = \SL(2,\C)/K, \\ 
{}\text{where }K = \SU(2), &\  \Conf(2) \cong \PGL(2,\C).
\end{split}
}
Similarly,
\deq{
\begin{split}
 S^1 = P^1(\R), &\ \,H^2 = \SL(2,\R)/K, \\
{}\text{where }K = \SO(2)), &\ 
 \Conf(1) \cong \PGL(2,\R). \end{split} 
 }
Now that everything is in purely algebraic language, depending only on a choice of field, we can simply make another choice! For example, we can take the $p$-adics:
\deq{
\begin{split}
 \partial T_p = P^1(\Q_p), &\ \,T_p = \SL(2,\Q_p)/K \\
 {}\text{where } K = \SL(2,\Z_p)), &\ 
 \Conf_p \cong \PGL(2,\Q_p).
 \end{split}
}
Here $T_p$ (the $p$-adic maximally symmetric space) is the famous ``tree'' of Bruhat and Tits: it is a uniform infinite tree of valence $p+1$. As we hoped, $T_p$ is naturally a discrete space, equipped with a metric---the ideal setting for lattice field theory. And there is another natural ``round'' measure, the Patterson--Sullivan measure $d\mu_0(x)$ on~$P^1(\Q_p)$, which stands in the same relation to the Haar measure as the Fubini--Study metric does to the translation-invariant metric on~$\C$.

Now, it's off to the races:
One can try to understand or formulate the simplest pieces of the holographic dictionary, starting with free bulk scalar fields propagating without backreaction on~$T_p$, just as in Witten's paper~\cite{HMSS,Gubs1}.
All of the necessary facts---compatible group actions on bulk and boundary, a bulk Klein-Gordon equation with a well-posed Dirichlet problem at infinity---are available; in fact, scalar fields on~$T_p$ were thoroughly studied in the context of the $p$-adic string theory (see, for example, \cite{Zabrodin}). 
{
The Klein--Gordon equation is 
\deq{
\Laplace \phi 
 = \sum_{v'\sim v} \left(\phi(v')-\phi(v)\right) = m^2 \phi , 
 }
and it admits a bulk-to-boundary Green's function solution, analogous to~$K(x)$:
 \deq{
  K (v) = p^{\kappa \langle v,x \rangle}.
}
This allows one to reconstruct the unique solution extending a given boundary condition, according to the formula 
\deq[reconstruct]{
\phi(v) = \frac{p}{p+1} \int_{\Q_p} d\mu_0(x)\, \phi_0(x) p^{\langle v,x \rangle} .
}
Here $\langle v,x \rangle$ is the distance from $v$ to the boundary point~$x$, regularized to be zero at the (arbitrary) center vertex~$C$ of the tree. $\kappa$ is a parameter; a moment's thought shows that the pair of~$\kappa$ and~$x$ is identical data to the wavevector $\mathbf{k}$ of an ordinary plane wave, corresponding to its magnitude and direction respectively, and that~$\kappa \langle v,x \rangle$ is a precise analogue of~$\mathbf{k} \cdot \mathbf{x}$.
The corresponding mass eigenvalue is
\deq{
 m_\kappa^2 = p^\kappa + p^{1-\kappa} - (p+1). 
 }
Thus, the BF bound is $m^2_\kappa \geq - (\sqrt{p}-1)^2$.
Just as in the normal case, there are solutions in the bulk whose mass-squared is  negative (but bounded from below)!
Bulk fields of mass $m_\kappa$ couple to boundary operators of conformal dimension~$\kappa$ (for a study of $p$-adic CFT, see e.g.~Melzer~\cite{Melzer}):
\deq{
\langle \mathscr{O}_\kappa(x) \mathscr{O}_\kappa(y) \rangle
\sim
\frac{1}{\left| x-y \right|_p^{2\kappa}}.
}

Pleasingly, some intuitive features of ordinary AdS/CFT---for example, that the radial coordinate represents a scale in the boundary theory---are present even more sharply in the $p$-adic case.
For instance, if the boundary field $\phi_0$ is a single mode (something like an additive character of~$\Q_p$, or one of the $p$-adic wavelets of Kozyrev~\cite{Kozyrev}), it stops contributing to the reconstruction of bulk physics abruptly, at a height determined by its wavelength!

To be precise, we can take the boundary condition to be of the form
\deq{
\phi_0(x) = \begin{cases} \exp 2\pi i \{ kx\}, & x \in \Z_p, \\
0 & x \not\in \Z_p.
\end{cases}
}
Let $v$ be a vertex in the branch of the tree above~$\Z_p$, at a depth~$\ell$ such that $0\leq \ell \leq -\ord_p(k)-1$. Then the  bulk function $\phi(v)$, reconstructed according to the prescription~\eqref{reconstruct}, is zero. The standard intuition that moving along the radial coordinate corresponds to a renormalization group scale in the boundary theory is thus confirmed.

\section{Analogues of gravity: edge length dynamics}

Since a crucial ingredient of ordinary AdS/CFT is the fact that the bulk theory includes gravity, it is natural to wonder what an analogue of gravity for the tree would look like. There is no completely satisfactory answer yet, but I will briefly describe one attempt to construct a theory for metric degrees of freedom~\cite{edgelength}.

The only obvious metric-type data in the bulk space $T_p$ are the edge lengths. Unlike an actual metric on a manifold, they don't have any obvious gauge freedom (since they are geodesic lengths and hence physical), but they do couple naturally to matter through the Laplacian. They are naturally set to a uniform value by the construction of the tree as a maximally symmetric space; when  they are deformed to a nonconstant configuration, the  group action on the tree is broken, but of course this is analogous to the normal case.


So it is natural to try allowing the edge lengths to fluctuate dynamically. Since the simplest possible coupling to matter is obvious, it remains only to come up with a sensible action principle for the edge lengths, which is some kind of analogue of the Einstein--Hilbert action for the metric.
A plausible choice of action can be defined using a notion of \emph{Ricci curvature} for graphs~\cite{Yau,Ollivier}, which has been studied in the mathematics literature. 

Since there are no loops in a tree-like graph and therefore no ``plaquette''-type operators like those studied in lattice gauge theory, it does not seem possible to define a curvature as the commutator of covariant derivatives. 
Rather, the graph Ricci curvature is defined in a global fashion, by computing the rate of change in the average (Wasserstein) distance between two separated heat kernels at $t=0$. 
 On a tree-like graph, it reduces to the following function on edges of the graph:
\begin{align*}
\kappa_{xy} &= \frac{b_{xy}}{d_x}\left( b_{xy} - \sum b_{xx_i}\right) + \frac{b_{xy}}{d_y}\left( b_{xy} - \sum b_{yy_i}\right)  \\
&\equiv \kappa_{x\rightarrow y} + \kappa_{y \rightarrow x}.
\end{align*}

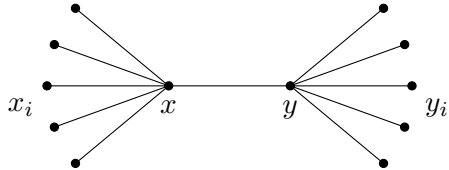
\begin{center}
\begin{tikzpicture}[scale=0.8]
\tikzstyle{vertex}=[draw,scale=0.3,fill=black,circle]
\draw (0,0) node[vertex,label=below:{$x$}]{} -- (2,0) node[vertex,label=below:{$y$}]{};
\draw (0,0) -- ++(140:2) node[vertex]{};
\draw (0,0) -- ++(160:2) node[vertex]{};
\draw (0,0) -- ++(180:2) node[vertex,label=below left:{$x_i$}]{};
\draw (0,0) -- ++(200:2) node[vertex]{};
\draw (0,0) -- ++(220:2) node[vertex]{};
\draw (2,0) -- ++(-40:2) node[vertex]{};
\draw (2,0) -- ++(-20:2) node[vertex]{};
\draw (2,0) -- ++(0:2) node[vertex,label=below right:{$y_i$}]{};
\draw (2,0) -- ++(20:2) node[vertex]{};
\draw (2,0) -- ++(400:2) node[vertex]{};
\end{tikzpicture}
\end{center}
Here $b$ denotes an inverse edge length, and $d_x = \sum\limits_{x'\sim x} b_{xx'}^2$.
As a straightforward sanity check, the Bruhat--Tits tree has constant negative curvature:
\deq{
\kappa_{xy} = -2 \frac{q-1}{q+1}.
}

So it seems reasonable to use $\kappa$ to write an analogue of the Einstein--Hilbert term.
The resulting action for a massive scalar field minimally coupled to edge length fluctuations, looks as follows:
\deq{
S = \sum_e \left(  (\kappa_e - 2 \Lambda) + \frac{b_e^2}{2} |\delta_e\phi|^2 \right) + \sum_v \frac{1}{2} m^2 \phi_v^2.
}

Now, in light of the holographic ansatz~\eqref{ansatz}, one is interested in the on-shell value of the classical action. But this is actually infinite for the solution corresponding to a uniform tree!
Just like in ordinary gravity, we need to regularize by a suitable choice of the constant term~$\Lambda$, and must also include a suitable boundary term (the analogue of the Gibbons--Hawking--York term) to ensure that the action makes sense when truncated to a finite region.

Let $\Gamma \subset T_p$ be a finite, connected subgraph, such that all vertices of~$\Gamma$ have valence either $p+1$ or~1. $\partial \Gamma$ is then the set of univalent vertices.
Then, one can consider truncating the action to~$\Gamma$: the action must be modified to
\deq{
S_\Gamma = S_\text{EH} + S_\text{bdy} 
= \sum_{e\in\Gamma} (\kappa_e - 2 \Lambda)    + \sum_{x \in \partial \Gamma} \ell_x,
}
where 
\deq{
\ell_x = K + \sum_{\substack{y\sim x \\ y \notin \Gamma}} \kappa_{x\rightarrow y}.
}
There are now two undetermined constants, $\Lambda$ and~$K$, in it; one constraint ensures that $S$ remains finite as $\Gamma \rightarrow T_p$:
\deq{
K 
= \frac{2q}{q-1}  \Lambda + q.  
}
When this constraint is satisfied, the on-shell action is
\deq{
S_\text{cl} = (2-2g) \left( 1 + \frac{q+1}{q-1}\Lambda \right).
}
Here $g$ is the genus (the result holds for the tree of any Mumford curve/higher-genus black hole). So the on-shell action is in fact \emph{topological!}

This suggests that this model is, perhaps, more like dilaton gravity in two dimensions than honest Einstein gravity.
Further evidence for that interpretation is provided by holographically computing correlation functions of the operator $T$ dual to edge-length fluctuations.
While the two-point functions are as one would expect for a massless bulk field,
 the three-point function actually vanishes identically, at least up to possible contact terms!

Just for the sake of completeness, here are some of those results more explicitly. The two point function of~$T$ is
\deq{
\langle T(z_1) T(z_2)  \rangle
= \frac{p^n}{4} \frac{\zeta(2n)}{\zeta(n)^2} \frac{1}{|z_{12}|^{2n}}.
}
For an operator $\mathscr{O}$ of dimension~$\kappa$,
\deq{
\langle T(z_1) \mathscr{O}(z_2) \mathscr{O}(z_3) \rangle
=  \frac{-\zeta(n) \zeta(2\kappa)}{\zeta(2\kappa - n) \zeta(-\kappa) \zeta(\kappa - n)} \frac{1}{|z_{12}|^{n}|z_{13}|^{n}|z_{23}|^{2\kappa - n}}.
}
And finally, 
\deq{
\langle T(z_1) T(z_2) T(z_3) \rangle = 0!
}
The reader is referred to~\cite{edgelength} for more details.

\section{Outlook}

Throughout this story, there is a long list of excellent questions, unclear points, and other things that remain (at least to the present author) opaque. We hope to continue to address some of these questions in future work, and hope that others will be motivated to take up the challenge as well.

For a start, there is an enormous amount of literature having to do with statistical mechanics models or spin systems defined on similar trees~\cite{Baxter-book}. It would be interesting to find any connection between such models and lattice field theories defined on trees, or $p$-adic boundary field theories. 

Based on the earlier discussion of the function field analogy, the reader may have wondered if a similar story is possible for local fields of finite characteristic (i.e., fields like $\FF{q}$ formal Laurent series). It would be interesting to think about such function field analogues in more detail, as well as about ad\`elic versions of the story. Furthermore, it is perhaps natural to wonder if there is any natural notion of field theory associated to the global object. While this is somewhat wild speculation, thinking about the function field case might be an easier and illustrative setting in which to try and define global notions.

On a somewhat more down-to-earth note, one also still lacks a clear candidate for a pair of dual theories, analogous to the pair provided by string theory in the normal case (although some ground-up steps in this direction have been taken~\cite{GubserParikh}). One part of the problem is that no $p$-adic theory nearly as complicated as maximally supersymmetric Yang--Mills seems to be readily available; one mostly has free theories, or theories of scalar fields with simple polynomial interactions.

 In the years since the AdS/CFT correspondence was first proposed, work has appeared that proposes versions of the holographic correspondence for simpler boundary theories, which are dual to correspondingly more complicated bulk gravitational theories; the $O(N)$ model, in particular, is meant to be related to a higher-spin gravity theory, as studied by Vasiliev~\cite{Vasiliev-grav, KlebanovPolyakov, Vasiliev}.
 Relatedly, it has been proposed~\cite{Razamat} that holographic duals of (Archimedean) bilocal free field theories can be constructed by using the exact renormalization group equations due to Polchinski. It would be very interesting to find such a construction for $p$-adic theories; we hope to report progress on this soon.

Another route of attack on this problem would involve coming up with more honest analogues of fields of nonzero spin. This would hopefully allow one to begin talking about gauge interactions, as well as more full-fledged analogues of gravity. 
The difficulty in doing this can be seen in several ways: for example, in the inability to construct any ``plaquette''-type operators as discussed above, or in the difficulty of defining a good notion of conserved current in $p$-adic field theory. Some vague speculation about using multiplicative characters of the field is given in~\cite{HMSS}, but this remains largely uncharted territory.

It would also be interesting to make a connection back to the subject of ``tensor network'' models, which are the other widespread class of discrete toy systems that are meant to relate to holography. Indeed, these models were part of the original motivation for the study of holography over the $p$-adics. In such a model, the bulk path integral is typically represented by the composition of many tensors or linear maps, acting between local (finite-dimensional) Hilbert spaces. One class of such tensors (``perfect'' tensors or quantum error-correcting codes) are naturally associated to the algebraic data of the tree, in particular to the geometry of algebraic curves over finite fields; such codes are reviewed in~\cite{HMSS}.

It is worth mentioning that there is, at least in a schematic sense, a natural tensor network (albeit with infinite-dimensional local Hilbert spaces) which \emph{is} the path integral of the bulk field. Given a region $\Gamma\subset T_p$, the Hilbert space is supposed to be constructed as the space of wave-functionals over the space of boundary conditions of fields; that is, for a single real scalar, it is (at least roughly) a tensor product of copies of $L^2(\R)$, one for each vertex $v \in \partial \Gamma$. The wave-functional, or time evolution matrix, is then determined by doing the path integral over the vertices in the interior of~$\Gamma$ with prescribed boundary conditions. Evolution ``across one vertex'' thus corresponds to the linear operator from $L^2(\R)$ to~$L^2(\R)^{\otimes p}$, with integral kernel 
\deq{
K(x;x_\nu) = 
\exp\left( - \sum_\nu (x - x_\nu)^2 - \mu x^2 
\right). 
}
Is there a precise sense in which this tensor network exhibits quantum error-correction properties? Does it admit natural truncations to a low-energy approximation, in which the local Hilbert spaces are finite-dimensional? If so, is there any connection of the resulting tensors to the algebrogeometric codes mentioned before? Or is there any other sort of natural path-integral model or spin system on~$T_p$ in which algebrogeometric perfect tensors naturally appear? We plan to address these questions in future work~\cite{HMSS-to-appear}.

\subsection*{Acknowledgements}

The author is deeply grateful to the organizing committee of the sixth international conference on $p$-adic mathematical physics and its applications, for the invitation to speak and for an enormously productive and enjoyable week. Special thanks are owed to W.~Z\'u\~niga, for his unsurpassed hospitality, as well as to B.~Dragovich, for the kind invitation to contribute this article to the conference proceedings and for his efforts towards their publication. The author also thanks all of the participants of the conference, for numerous good questions and insight-laden conversation; I am grateful for everything I learned from you all.

None of this work would have been possible without the diligence and talent of my collaborators. I am so fortunate to have had the chance to work with you all. Thanks to each of you:
S.~Gubser, M.~Heydeman, C.~Jepsen, M.~Marcolli, S.~Parikh, B.~Stoica, B.~Trundy, and P.~Witaszczyk.
Gratitude is due also to the many friends and colleagues who have patiently listened to chatter about this work, provided opportunities for it to be spoken about, and offered invaluable conversation and feedback. Thanks to you all, and especially to 
T.~Dimofte, V.~Disarlo, R.~Eager, M.~Fluder, V.~Forini, J.~Keating, M.~Kim, P.~Longhi, T.~McKinney, B.~Pozzetti, M.~Salmhofer, A.~Schwarz,  J.~Walcher, A.~Waldron, D.~Xie, W.~Yan, S.-T.~Yau, and M.~Zidikis.

The author's work is supported in part by the Deutsche Forschungsgemeinschaft, within the
framework of the Exzellenzinitiative an der Universit{\"a}t Heidelberg;
he wishes to also express thanks to 
Roost Coffee \&\ Market, where a large fraction of this manuscript was prepared.

\pagebreak[1]

\bibliographystyle{chicago}
\bibliography{proceedings.bib}

\end{document}